# 2D Lyndon Words and Applications

Shoshana Marcus[*]    Dina Sokol [†]

**Abstract**

A Lyndon word is a primitive string which is lexicographically smallest among cyclic permutations of its characters. Lyndon words are used for constructing bases in free Lie algebras, constructing de Bruijn sequences, finding the lexicographically smallest or largest substring in a string, and succinct suffix-prefix matching of highly periodic strings. In this paper, we extend the concept of the Lyndon word to two dimensions. We introduce the 2D Lyndon word and use it to capture 2D horizontal periodicity of a matrix in which each row is highly periodic, and to efficiently solve 2D horizontal suffix-prefix matching among a set of patterns. This yields a succinct and efficient algorithm for 2D dictionary matching.

We present several algorithms that compute the 2D Lyndon word that represents a matrix. The final algorithm achieves linear time complexity even when the least common multiple of the periods of the rows is exponential in the matrix width.

## 1 Introduction

Two strings are *conjugate* if they differ only by a cyclic permutation of their characters. A *Lyndon word* is a primitive string which is the smallest of its conjugates for the alphabetic ordering [11]. For example, abba and aabb are conjugate; aabb is a Lyndon word, while abba is not. Lyndon words are useful for constructing bases in free Lie algebras [12], constructing de Bruijn sequences [6], computing the lexicographically smallest or largest substring in a string [4], and succinct suffix-prefix matching of highly periodic strings [13].

---

[*]Simons Center for Quantitative Biology, Cold Spring Harbor Laboratory, 1 Bungtown Road, Cold Spring Harbor, NY, 11724. email: smarcus@cshl.edu.

[†]Department of Computer and Information Science, Brooklyn College of the City University of New York, 2900 Bedford Avenue, Brooklyn, N.Y. 11210. email: sokol@sci.brooklyn.cuny.edu.

A string $S$ is *periodic* if it can be expressed as $u^j u'$ where $u'$ is a proper prefix of $u$, and $j \geq 2$. When $u$ is primitive, i.e., it is not a power of another string, we call it "the period" of $S$. Depending on the context, we use the term *period* to refer to either $u$ or $|u|$. A string is periodic if it can be superimposed upon itself before its midpoint.

*Lyndon word naming* classifies highly periodic strings by the conjugacy of their periods and uses the Lyndon word as the class representative. Once Lyndon word naming has been performed, a string can be represented by the name of its period's class and its *LWpos*, the position at which the Lyndon word first occurs in the string. For example, the strings $T_1$ = abbaabbaabbaabbaab and $T_2$ = aabbaabbaabbaabbaa are in the same class and the class representative is aabb. *LWpos* of $T_1$ is 3 while *LWpos* of $T_2$ is 0 since it begins with the Lyndon word that represents its period [13].

In this paper, we extend the concept of the Lyndon word to two dimensions. The *2D Lyndon word* is a succinct representation of the Lyndon word that is conjugate to each row combined with the relative alignments of the Lyndon words among the matrix rows. We introduce a new classification scheme based on the 2D Lyndon word for matrices whose rows are highly periodic. This new classification scheme captures the horizontal suffix-prefix matches among a set of matrices. We deal with matrices whose rows are highly periodic since a non-periodic pattern row would allow us to quickly narrow down the possible pattern occurrences in a text yielding much fewer possibilities of overlap.

It is straightforward to compute the 2D Lyndon word that represents a matrix, however, the naive computation takes exponential time for certain inputs. We use modular arithmetic to develop an extremely efficient algorithm that does not need to process the actual matrix. We present several algorithms to compute the 2D Lyndon word that represents a matrix. Their time complexities are summarized in Table 1. The input to these algorithms is the output of Lyndon word naming on the rows, i.e., the period size of each row, and the offset of the first Lyndon word occurrence in each row. $LCM_m$ denotes the least common multiple of the periods of all rows.

The classification technique that we introduce is a new way of capturing horizontal 2D periodicity. Amir and Benson introduced the concept of 2D periodicity [1, 2] and it serves as the basis for an efficient 2D pattern matching algorithm [3]. However, their approach to 2D periodicity is not suitable for multiple pattern matching, which requires suffix-prefix matching between pairs of different patterns. The *all-pairs suffix-prefix matching problem* is the problem of finding, for any pair of strings in a given set, the longest suffix of one which is a prefix of the other. Lyndon word naming is an efficient tool to identify suffix-prefix matches between highly periodic strings. 2D Lyndon word naming is equally meaningful for horizontal suffix-prefix matching in matrices, resulting in efficient dictionary matching. Both one and two dimensional Lyndon word naming have the adddi-tional benefit over other algorithms, e.g. [7, 8, 14], of being online and of using very little working space.

Table 1: A summary of the algorithms presented in this paper to compute the 2D Lyndon word that represents an $m \times m$ matrix.

|  | Best Time Complexity | Worst Time Complexity | Described In |
|---|---|---|---|
| Naive Algorithm | $O(m \cdot LCM_m)$ | $O(m \cdot LCM_m)$ | Section 3 |
| Algorithm 1 | $O(m \log^2 m + LCM_m)$ | $O(m \log^2 m + (LCM_m + m)\frac{m}{\log m})$ | Section 3 |
| Algorithm 2 | $O(m \log^2 m)$ | $O(m \log^2 m + \frac{m^2}{\log m})$ | Section 4 |

The remainder of this paper is organized as follows. We begin by defining the 2D Lyndon word in Section 2. In Section 3, we present an algorithm that calculates the 2D Lyndon word directly from the actual matrix. In Section 4, we present a more efficient algorithm that calculates the 2D Lyndon word using modular arithmetic. In Section 5, we apply this technique and show how it is useful for several applications. We conclude with a summary in Section 6.

## 2 Main Idea

### 2.1 Definition of 2D Lyndon word

We say that a matrix $M$ has a *horizontal prefix* (resp. suffix) $U$ if $U$ is an initial (resp. ending) sequence of contiguous columns in $M$. A matrix is *horizontally primitive*, or *h-primitive*, if it cannot be written in the form $U^i$ for any horizontal prefix $U$ and integer $i > 1$.

**Definition 2.1** *Two matrices, $M_1$ and $M_2$, are* horizontal 2D conjugate *if $M_1 = UV$, $M_2 = VU$ for some horizontal prefix $U$ and horizontal suffix $V$ of $M_1$.*

We say that two matrices are *horizontal 2D conjugate* if they differ only by a cyclic permutation of entire columns. When it is clear from the context, we simply use the word conjugate to refer to horizontal 2D conjugate. We show in Lemma 1 that horizontal 2D conjugacy defines an equivalence relation.

**Lemma 1** *Horizontal 2D conjugacy is an equivalence relation among matrices.*

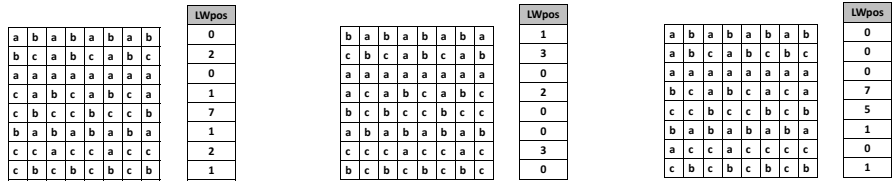

Figure 1: Matrices that are horizontal 2D conjugate, along with each *LWpos* array. In its first conjugate, the matrix is shifted right by one column and in its second conjugate it is shifted left by two columns. The matrix on the right is a 2D Lyndon word.

**Proof:** In one-dimension, strings $x$ and $y$ are defined as conjugate if $x = uv$, $y = vu$ for some strings $u$, $v$. It is easy to show that one-dimensional conjugacy is reflexive, symmetric, and transitive, and thus is an equivalence relation [11]. We can reduce horizontal 2D conjugacy to 1D conjugacy by viewing each entire column as a metacharacter in the cyclic permutation, thus yielding an equivalence relation. ∎

We represent each horizontal 2D conjugate as a sequence $c_1, c_2, \ldots, c_m$ where $c_i$, $1 \leq i \leq m$, represents the minimum number of characters that need to be cyclically permuted in row $i$ to obtain a 1D Lyndon word. For example, if row $i$ is the string $T = uv$, $u$ is a prefix of $T$, $v$ is a suffix of $T$, and $vu$ is a Lyndon word, $c_i = |u|$. We refer to the sequence of a conjugate as the *LWpos* array since it is essentially an array of Lyndon word positions in the matrix.

We order horizontal 2D conjugates by comparing their *LWpos* arrays. Three matrices that are horizontal 2D conjugate are depicted in Fig. 1, along with their *LWpos* arrays. The order of the matrices in ascending order is: the matrix on the right, the matrix on the left, and then the matrix in the center.

**Definition 2.2** *A* 2D Lyndon word *is an h-primitive matrix that is the smallest of its horizontal 2D conjugates for the numerical ordering of* LWpos *arrays.*

The 2D Lyndon word is defined for all h-primitive matrices. It is a succinct representation of the Lyndon word that is conjugate to each row, combined with the relative alignments of the Lyndon words among the matrix rows. In one dimension, Lyndon word naming provides a classification of highly periodic strings based upon the Lyndon word of the *period* of the string, which is by definition primitive. Analogous to this in two dimensions, we compute the representative 2D Lyndon word of the 2D horizontal period, or more specifically, the h-primitive *LCM-matrix*, as described in the next subsection.

Figure 2: A matrix with its *LCM-matrix* highlighted. The periods of the rows are of length 1, 2 and 3. $LCM_m = 6$, yielding an *LCM-matrix* that is 6 columns wide.

## 2.2 Classification Scheme

In this section, we present a new classification scheme for matrices whose rows are all highly periodic (i.e. for a matrix of width $m$, all rows have period $\leq m/4$). In a matrix whose rows are all highly periodic, the columns may also repeat at regular intervals. This matrix-wide repetition is at a distance of the lowest common multiple (LCM) of the periods of the rows, as we prove in Lemma 2. If the columns repeat, we focus on a submatrix that spans the first $LCM_m$ columns of the original matrix. If $LCM_m$ spans up to $m/2$ columns, then $LCM_m$ is in fact a horizontal period of the matrix. If the width of the matrix is smaller than $LCM_m$, then we can enlarge the matrix to $LCM_m$ columns by extending the period in each row. We refer to the (possibly enlarged) matrix of width $LCM_m$ as the *LCM-matrix* of the original matrix. Fig. 2 shows a matrix with its *LCM-matrix* highlighted. We use a matrix with a small $LCM_m$ in the example to illustrate the definitions, however, all the definitions and algorithms apply to a matrix with a large $LCM_m$ as well.

**Lemma 2** *In a matrix with $m$ rows, each of which is a periodic string, the columns repeat every $LCM_m$ columns, where $LCM_m$ denotes the least common multiple of the periods of all rows.*

**Proof:** Every row repeats in columns that are multiples of its period. $LCM_m$ is a multiple of every row's period. Since every row repeats at $LCM_m$ columns, the entire matrix repeats at $LCM_m$ columns. ∎

It follows from Lemma 2 that each of the $LCM_m$ conjugates of an *LCM-matrix* has a distinct *LWpos* array. The key property of an *LCM-matrix* is that horizontal 2D conjugacy preserves row periodicity. We prove this in Lemma 3 by showing that a cyclic permutation of the columns in an *LCM-matrix* results in a cyclic permutation of each row's period.

**Lemma 3** *Two* LCM-matrices *that are horizontal 2D conjugate have periods in corresponding rows that are 1D conjugate.*

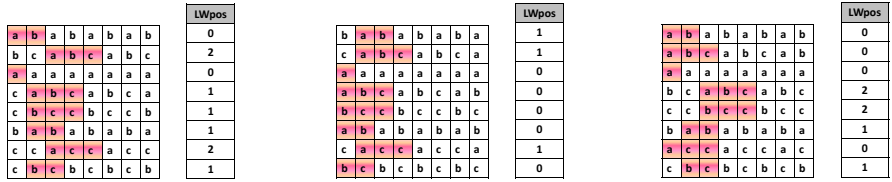

Figure 3: Matrices whose *LCM-matrices* are horizontal 2D conjugate, along with each *LCM-matrix*'s *LWpos* array. These matrices are not horizontal 2D conjugate. In each of these matrices, $LCM_m = 6$. The first occurrence of the Lyndon word in each row is highlighted. The first matrix appears in Fig. 2, where the focus is on its *LCM-matrix*. In its first conjugate, the *LCM-matrix* is shifted left by one column and in its second conjugate it is shifted left by two columns.

**Proof:** A cyclic permutation of columns in an *LCM-matrix* moves columns to a distance of $LCM_m$. By Lemma 2, the columns of the matrix repeat every $LCM_m$ columns. Therefore, a cyclic permutation of columns in an *LCM-matrix* is equivalent to removing several columns from one end of the matrix and extending the period in each row at the other end of the matrix. Thus, a cyclic permutation of columns in the *LCM-matrix* maintains the period of each row, up to a cyclic permutation of each period. ∎

In our new classification scheme, each equivalence class consists of matrices whose *LCM-matrices* are horizontal 2D conjugate. We use the 2D Lyndon word in each class as the class representative, in a similar manner to the 1D equivalence relation that uses the *Lyndon word* as the class representative. Three matrices whose *LCM-matrices* are horizontal 2D conjugate are shown in Fig. 3, along with their *LWpos* arrays. To classify matrix $M$ as belonging to exactly one horizontal 2D conjugacy class, we compute the conjugate of $M$'s *LCM-matrix* that is a 2D Lyndon word. This classification allows us to represent a matrix succinctly, with a constant number of 1D arrays. At the same time, this representation allows us to quickly answer horizontal suffix-prefix queries on a set of classified matrices. In the following two sections, we present several algorithms for computing the 2D Lyndon word that represents a given matrix.

## 3  Simple Algorithm for Computing 2D Lyndon word

In this section we develop an intuitive algorithm that efficiently computes the 2D Lyndon word to represent an $m \times m$ matrix whose rows are highly periodic. We present algorithms that are run after Lyndon word naming has been performed on each row of the matrix. That is, the input to each of these algorithms is an $m \times m$ matrix represented by 3 arrays of size $m$, the 1D Lyndon word names for each row, the period size of each row, and an *LWpos* array. Lyndon word naming of the matrix rows takes linear $O(m^2)$ time [13]. 1D Lyndon word naming was designed for highly periodic strings, with periods $\leq m/4$. As in [13], these ideas can be extended from squares to rectangles of uniform size in at least one dimension.

| 0 | 1 | 0 | 1 | 0 | 1 |
|---|---|---|---|---|---|
| 2 | 1 | 0 | 2 | 1 | 0 |
| 0 | 0 | 0 | 0 | 0 | 0 |
| 1 | 0 | 2 | 1 | 0 | 2 |
| 1 | 0 | 2 | 1 | 0 | 2 |
| 1 | 0 | 1 | 0 | 1 | 0 |
| 2 | 1 | 0 | 2 | 1 | 0 |
| 1 | 0 | 1 | 0 | 1 | 0 |

Figure 4: The set of *LWpos* arrays for the conjugates of an *LCM-matrix*. Each column in this table contains the *LWpos* array of the conjugate that begins with that column. The column that begins the 2D Lyndon word is highlighted. This *LCM-matrix* corresponds to the matrix in Fig. 2 and the leftmost matrix in Fig. 3.

We have already seen that each conjugate of an *LCM-matrix* can be obtained by a cyclic permutation of columns in the *LCM-matrix*. As a result, computing the 2D Lyndon word that represents a matrix is a search for the cyclic permutation of its *LCM-matrix* at which the *LWpos* array is smallest.

We can naively compute the 2D Lyndon word that represents a matrix by computing the *LWpos* array for each conjugate of its *LCM-matrix* and then finding the minimum sequence. We show in Lemma 4 that the conjugates can be obtained by shifting the matrix rows. Thus, we can generate each conjugate's *LWpos* array from the matrix's *LWpos* array combined with the periods of the rows and then select the minimum *LWpos* array in this set. Since the *LCM-matrix* has $LCM_m$ conjugates to consider (by Lemma 2), the naive algorithm runs in time proportional to the size of the *LCM-matrix*, $O(m \cdot LCM_m)$ time. Fig. 4 shows the set of *LWpos* arrays for the conjugates of the *LCM-matrix* depicted in Fig. 2. Each column represents the *LWpos* array of the conjugate that begins with that column. The columns of this table are compared from top-down and the minimum is selected as the 2D Lyndon word. In this example, the conjugate that begins with the third column is the 2D Lyndon word that represents the matrix depicted in Fig. 2.

**Lemma 4** *Two matrices have* LCM-matrices *that are* horizontal 2D conjugate *iff the* LWpos *entries for each row are shifted by* $C \pmod{period[i]}$, *where C is an integer and period[i] is the period size of row i.*

**Proof:** Suppose matrices $M_1$ and $M_2$ have *LCM-matrices* that are horizontal 2D conjugate. The *LCM-matrix* of $M_1$ is obtained from $M_2$'s by a cyclic permutation of columns. Thus, the corresponding rows in the *LCM-matrices* are each cyclically shifted by $C$ characters. The *LWpos* of row $i$ can range from 0 to *period*$[i]$. A shift of $C$ characters translates to a shift of $C \pmod{period[i]}$ on row $i$, by Lemma 3. Similarly, if we know that the shift of each row $i$ is $C \pmod{period[i]}$, $0 \leq i < m$, the *LCM-matrices* must be horizontal 2D conjugate. ∎

We improve on the naive algorithm and present an $O(m + LCM_m)$ time algorithm for calculating the 2D Lyndon word that represents an $m \times m$ matrix. This procedure is delineated in Algorithm 1 and described in the following paragraphs.

We can systematically compute the numerically smallest *LWpos* array among the conjugates of the *LCM-matrix* without actually generating the complete *LWpos* arrays. The computation is incremental and considers one row at a time. Initially, before we examine the first row, all columns of the *LCM-matrix* are potentially the beginning of the 2D Lyndon word. As we proceed through the rows, we discard columns that cannot be the beginning of the 2D Lyndon word. Once a column is discarded, it is never considered again.

In our example in Fig. 4, we begin by eliminating the second, fourth, and sixth columns since they represent conjugates in which the first row's Lyndon word offset is 1, which is larger than 0. When we get to the second row, we focus on the first, third, and fifth columns and ignore the others. The Lyndon word offsets in these columns are 2, 0, and 1, with 0 the minimum of these values. Thus, we select the third column as the beginning of the 2D Lyndon word since the corresponding *LWpos* array begins with 00 and all the other conjugates begin with larger values. If there would be several columns that begin with 00, this process would continue until only one column remains.

In general, we begin by eliminating all but the columns at which the Lyndon word of the first row begins. Suppose the first *LWpos* entry is $z$ and the period of the first row is $u$. Columns $z, z+u, z+2u, \ldots$ are the only columns at which the 2D Lyndon word can begin; the other columns are immediately eliminated.

Subsequently, for each row $i$ there are two possibilities. The first possibility is that the period of row $i$ is a factor of the least common multiple of the periods of the first $i-1$ rows, which we denote by $LCM[i-1]$. In this case, the Lyndon word offset is identical in all remaining columns. We calculate the *LWpos* entry without eliminating any columns. The other possibility is that the period of row $i$ is not a factor of $LCM[i-1]$, i.e., $LCM[i]$ is larger than $LCM[i-1]$. In this case, we calculate the *LWpos* value in each remaining column, select the minimum, and update $z$ to be the first column that attains this minimum value. Then, columns

**Algorithm 1** Computing a 2D Lyndon Word
―――――――――――――――――――――――――――――――――――――――――
Input: $LWpos[1...m], period[1...m]$ for matrix $M$.
Output: 2D Lyndon word, $2D\_LW[1...m]$, and its shift $z$ (i.e. column number in *LCM-matrix* of $M$).

$2D\_LW[1] \leftarrow 0$
$z \leftarrow LWpos[1]$
{*LWpos*[1] is first column of shift 0}
{columns $z, z + period[1], z + 2 * period[1], \ldots$ can be 2D Lyndon word}
$LCM[1] \leftarrow period[1]$
**for** $i \leftarrow 2$ to $m$ **do**
   $GCD \leftarrow \gcd(LCM[i-1], period[i])$
   $LCM[i] \leftarrow LCM[i-1]*period[i]/GCD$
   **if** $LCM[i-1] \equiv 0 \pmod{period[i]}$ **then**
     {if period of row $i$ is a factor of cumulative LCM}
     $2D\_LW[i] \leftarrow (LWpos[i] - z) \pmod{period[i]}$
   **else**
     {$LCM[i] > LCM[i-1]$}
     *firstShift* $\leftarrow (LWpos[i] - z) \pmod{period[i]}$
     {shift $LWpos[i]$ to column $z$}
     $2D\_LW[i] \leftarrow \min((firstShift - x * LCM[i-1]) \pmod{period[i]})$
     {minimize over $x \geq 0$ such that $z + x * LCM[i-1] \leq LCM[m]$}
     $z += x * LCM[i-1]$
     {adjust $z$ by $x$ that minimizes shift in previous equation}
   **end if**
**end for**
―――――――――――――――――――――――――――――――――――――――――

$z, z + u, z + 2u, \ldots$, where $u = LCM[i]$, are the only columns at which the 2D Lyndon word can begin, since the columns of the first $i$ rows in the table of *LWpos* arrays recur every $LCM[i]$ columns, by Lemma 2. This process continues until the last row is reached and only one column remains, since the columns in an *LCM-matrix* are distinct.

**Lemma 5** *Let M be an $m \times m$ matrix and let $\alpha$ denote the time complexity of a single arithmetic operation on $LCM_m$ of the matrix and a second operand that is $\leq m$. Algorithm 1 computes the 2D Lyndon word that represents M in $O(m \log^2 m + (LCM_m + m)\alpha)$ time and uses $O(m \log m)$ bits of working space.*

**Proof:** The greatest common divisor of $LCM[i-1]$ and $period[i]$ can be computed in $O(\log^2 m)$ time since the Euclidean algorithm takes $O(\log^2 m)$ time to compute the greatest common divisor of two integers when the smaller operand is stored in $\log m$ bits [9], after the first modulus step that requires $O(\alpha)$ time. In this case, *period*[$i$] $\leq m/4$ is stored in $\log m$ bits and *LCM*[$i-1$] may be larger. Subsequently, the least common multiple of *LCM*[$i-1$] and *period*[$i$] is computed from their greatest common divisor in $O(\alpha)$ time. Over all rows, the total time spent on LCM computations is $O(m \log^2 m + m\alpha)$.

The *LCM-matrix* has $LCM_m$ distinct columns, by Lemma 2. Thus, Algorithm 1 begins with a set of $LCM_m$ columns at which the 2D Lyndon word can begin. As row $i$ is examined, the if statement in Algorithm 1 has two possibilities:
(i) Its period is a factor of *LCM*[$i-1$]: computation completes in $O(\alpha)$ time.
(ii) *LCM*[$i$] > *LCM*[$i-1$]: we examine the *LWpos* arrays for the conjugates beginning in several columns. The values are compared and all but the columns of minimal value are discarded. Since $LCM(x,y) > 2x$ where $x > y$, and $y$ is not a factor of $x$, at least half the possibilities are discarded, and we can charge the computation of *LWpos* values in row $i$ to the discarded columns. Over all rows, at most $LCM_m$ columns can be discarded. The computation of an *LWpos* value takes $O(\alpha)$ time.

Thus, the overall time complexity, aside from the LCM computation, is $O((LCM_m + m)\alpha)$. In terms of space, the representative 2D Lyndon word is stored in $O(m \log m)$ bits since it is an array of $m$ integers, each of which is between 0 and $m/4$. Along the way, the only extra information we store are the column number, $z$, and *LWpos* values for the active column and for the minimum in the preceding columns of the row. ∎

In the best case, $LCM_m$ is linear or polynomial in $m$, thus it can be stored in $O(\log m)$ bits and fits in one word of memory. Then, $\alpha = O(1)$, and the algorithm runs in $O(m \log^2 m + LCM_m)$ time. In the worst case, $LCM_m$ can grow exponentially, yet an upper bound of $3^m$ has been proven for the LCM of the numbers 1 through $m$ [5]. Thus, the least common multiples can always be stored in $O(m)$ bits and $\alpha$ is at most $O(m/\log m)$. Hence, the worst case running time of Algorithm 1 is $O(m \log^2 m + (LCM_m + m)\frac{m}{\log m})$.

Since Algorithm 1 requires exponential time with respect to the input size in the worst case, in the next section we present a different algorithm whose time complexity is dependent on the number of bits needed to store $LCM_m$, yielding a worst case linear time algorithm for computing a 2D Lyndon word. We compare the time complexities of these algorithms in Table 1.

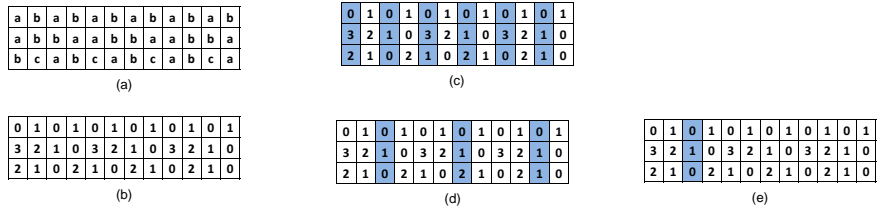

Figure 5: (a) An *LCM-matrix*. (b) Its table of *LWpos* arrays in which each column contains the *LWpos* array of the conjugate that begins with that column. (c)-(e) The computation of the 2D Lyndon word that represents the matrix. The columns that remain after each iteration of the algorithm are highlighted. (c) After examining the first row, the columns beginning with 0 remain. (d) After examining the first two rows, the columns beginning with 01 remain. (e) The only remaining column is the 2D Lyndon word.

# 4 Computation of 2D Lyndon word by Modular Arithmetic

In this section we derive a more efficient algorithm to compute the 2D Lyndon word that represents a matrix. The naive algorithm generates the *LWpos* array for each conjugate of the *LCM-matrix*, as shown in Fig. 5(b), and selects the minimum. Algorithm 1 only partially computes the *LWpos* arrays and narrows in on the column at which the 2D Lyndon word begins, as computation proceeds through the rows of the matrix. This is depicted in Fig. 5(c),(d), and (e). In this section, we present Algorithm 2, which uses modular arithmetic to directly compute each *LWpos* entry of the 2D Lyndon word, avoiding the comparison of any *LWpos* entries of the *LCM-matrix*'s conjugates. We show a reduction of the computation of a representative 2D Lyndon word to an algebraic problem that is solved with modular arithmetic.

In Algorithm 1, we obtain the representative 2D Lyndon word by computing *LWpos* values for some of the conjugates of the *LCM-matrix* and then selecting the minimum value. We transform this to a sequence for each row $i$,

$$S_i[x] = (f - \ell X) \pmod{p}$$

where $\ell$ is $LCM[i-1]$, $p$ is $period[i]$, and $f$ is the first column for which we consider an *LWpos* entry for row $i$. The objective of Algorithm 1 is to find the minimum value in $S_i$ and the value of $X$ at its first occurrence. We use properties of modular arithmetic to solve this problem.

**Algorithm 2** Computing a 2D Lyndon Word More Efficiently
---
Input: $LWpos[1...m], period[1...m]$ for matrix $M$.
Output: 2D Lyndon word, $2D\_LW[1...m]$, and its shift $z$ (i.e. column number in *LCM-matrix* of $M$).

$2D\_LW[1] \leftarrow 0$
$z \leftarrow LWpos[1]$
{*LWpos*[1] is first column of shift 0}
$LCM[1] \leftarrow period[1]$
**for** $i \leftarrow 2$ to $m$ **do**
   $GCD \leftarrow \gcd(LCM[i-1], period[i])$
   $\ell \leftarrow LCM[i-1]/GCD$
   $p \leftarrow period[i]/GCD$
   $\ell Inv \leftarrow$ inverse of $\ell \pmod{p}$
   $LCM[i] \leftarrow \ell * period[i]$
   *firstShift* $\leftarrow (LWpos[i]-z) \pmod{period[i]}$
   {shift *LWpos*[$i$] to $z$}
   *divFirstShift* $\leftarrow \lfloor \textit{firstShift} / GCD \rfloor$
   $x \leftarrow (\ell Inv * \textit{divFirstShift}) \pmod{p}$
   $2D\_LW[i] \leftarrow (\textit{firstShift} - x*LCM[i-1]) \pmod{period[i]}$
   $z+ = x*LCM[i-1]$
**end for**
---

**Definition 4.1** *[10] The* modular inverse *of an integer $\ell \pmod{p}$ is an integer $\ell^{-1}$ such that $\ell(\ell^{-1}) \equiv 1 \pmod{p}$. More simply, we refer to $\ell^{-1}$ as an inverse.*

The modular inverse of $\ell \pmod{p}$ exists iff $\gcd(\ell, p) = 1$. In other words, $\ell \pmod{p}$ has an inverse when $\ell$ and $p$ are *relatively prime* [10].

When $\ell$ and $p$ are relatively prime, 0 is the minimum value in the sequence and $\ell^{-1} \pmod{p}$ is the first position $x$ for which $S_i[x] = 1$. Multiplying $\ell^{-1}$ by the first value in the sequence, $\ell^{-1} * f \pmod{p}$, locates the first position $x$ such that $S_i[x] = 0$, the first minimum in the sequence.

When $\ell$ and $p$ are not relatively prime, the minimum value in $S_i$ may not be 0. We can convert $S_i$ to a sequence with a minimum of 0 by dividing both $\ell$ and $p$ by their greatest common divisor. Then 0 is surely in $S_i$ and we can use $\ell^{-1}$ to locate its first occurrence, as before. The process of computing the representative 2D Lyndon word by modular arithmetic is delineated in Algorithm 2.

**Lemma 6** *Let LWpos[1 ... m] and period[1 ... m] be the input to Algorithm 2. Let $LCM[0] = 1$ and $x[1] = LWpos[1]$. When Algorithm 2 completes, the column of the 2D Lyndon word in the* LCM-matrix $z = \sum_{i=1}^{m}(x[i]*LCM[i-1])$.

**Proof:** This summation directly follows from the pseudocode for Algorithm 2. $z$ is initialized to *LWpos*[1] on line 2, then for each $i$, a new $x$-value is calculated, and on the last line, $x * LCM[i-1]$ is added to $z$. ∎

**Lemma 7** *Let M be an $m \times m$ matrix and let $\alpha$ denote the time complexity of a single arithmetic operation on $LCM_m$ of the matrix and a second operand that is $\leq m$. Algorithm 2 computes the 2D Lyndon word that represents M in $O(m \log^2 m + m\alpha)$ time and uses $O(m \log m)$ bits of working space.*

**Proof:** As we showed in the proof of Lemma 5, the Euclidean algorithm for the GCD of $LCM[i-1]$ and *period*[$i$] runs in $O(\log^2 m + \alpha)$ time. Subsequently, the LCM and modular inverse for each row are computed in $O(\alpha)$ time with basic arithmetic. These computations take $O(\log^2 m + m\alpha)$ time for the entire matrix.

Some of the variables in Algorithm 2 can be large integers whose sizes are related to $LCM_m$ and require $O(m)$ bits to store. All arithmetic operations in the algorithm that have one operand that is a large integer have a small second operand so they run in $O(\alpha)$ time. Once the GCD and LCM have been computed, the other steps take at most $O(\alpha)$ time per row, which is $O(m\alpha)$ time for the entire matrix.

As we showed in the proof of Lemma 5, the 2D Lyndon word that represents M is stored in $O(m \log m)$ bits of space. The intermediate results can all be stored in $O(m)$ bits of space. ∎

The best case is where $LCM_m$ is polynomial in $m$, so $\alpha = O(1)$ and Algorithm 2 runs in sublinear $O(m \log^2 m)$ time. The worst case is where $LCM_m = O(3^m)$, resulting in $\alpha = O(m/\log m)$, yielding worst case time complexity of $O(m \log^2 m + \frac{m^2}{\log m})$.

## 5 Applications

### 5.1 2D Periodicity

The classification technique that we introduce in this paper is a new perspective on horizontal 2D periodicity. When Amir and Benson introduced the concept of 2D periodicity [1, 2], they presented matrix periodicity as self-overlap that covers the center of the matrix. Our classification scheme is based on horizontal periodicity in a matrix. Just as Amir and Benson's 2D periodicity is the basis for an efficient 2D pattern matching algorithm [3], so is horizontal periodicity. Our new techniques have the benefit of being succinct since we do not need to store a 2D witness table for each pattern. Furthermore, our techniques generalize nicely to multiple patterns, as we show in the next two sections.

## 5.2 Suffix-Prefix Matching

The 2D Lyndon word naming technique contributes the first efficient tool for horizontal suffix-prefix matching in a set of matrices whose rows are all highly periodic. Two $m \times m$ matrices whose rows are all periodic, $M_1$ and $M_2$, can have a horizontal suffix-prefix match of $\geq m/2$ columns if the *LCM-matrices* of $M_1$ and $M_2$ are horizontal 2D conjugate. When two matrices are in the same class, the difference between the number of columns that are cyclically permuted in each *LCM-matrix* determines whether there is a horizontal suffix-prefix match, and if so, by how many columns. After linear time preprocessing classifies each matrix in a set, horizontal suffix-prefix queries between two matrices are answered in constant time. This algorithm is succinct since it uses only $O(km)$ extra space for input of size $O(km^2)$. It is online since matrices can be classified as they arrive and horizontal suffix-prefix matches can be announced at any time.

## 5.3 Succinct Dictionary Matching with the 2D Lyndon Word

The goal of *dictionary matching* is to search a text for all occurrences of any pattern from a given set of $d$ patterns. We focus on patterns whose rows are periodic with period $\leq m/4$ since that is the more challenging case for a succinct algorithm. A non-periodic pattern row would allow us to quickly narrow down the possible pattern occurrences in a text, yielding much fewer possibilities of overlap. Our new classification technique improves the succinct 2D dictionary matching algorithm of [13], which has a strict implicit assumption that the period of the first row of each pattern matches the horizontal period of the pattern. We work with $3m/2 \times 3m/2$ blocks of the text to conserve working space. For succinct dictionary matching on 2D data whose rows are highly periodic, Lyndon word naming uniquely names each row of the patterns and of the text [13]. Then we perform 1D dictionary matching on the linearized dictionary of patterns and small blocks of the linearized text to identify candidates for pattern occurrences.

The final stage of verification confirms that the Lyndon words are correctly aligned in corresponding pattern rows and text rows. Segments of the larger text need to be compared to many patterns simultaneously, however, we consider only one member of each class of patterns that can overlap in a text block of size $3m/2 \times 3m/2$, namely the representative 2D Lyndon word. When $LCM_m$ of the candidate patterns is polynomial in $m$, during pattern preprocessing, a compressed trie is constructed for the set of 2D Lyndon words that represent patterns with the same 1D pattern of names. Then, at text verification, we can compute the 2D Lyndon word to represent the text block beginning at each candidate row in $O(m)$ time, provided that we have access to the *GCD* values that were computed for the patterns. We then traverse a compressed trie of 2D Lyndon words to see if any pattern is consistent with the text block. In total, linear $O(m^2)$ time suffices to verify a $3m/2 \times 3m/2$ block of text.

However, when the least common multiple of the periods of the pattern rows (*LCM$_m$*) is exponential in $m$, the algorithm would result in $O(n^3)$ time for a text of size $n^2$. In this case, arithmetic operations that involve the *LCM* of the periods of the pattern rows cannot necessarily complete in $O(1)$ time. Thus, in this section we propose a more sophisticated algorithm for text verification that avoids these expensive computations. We show that partial computation of each submatrix's representative 2D Lyndon word suffices to verify pattern occurrences, yielding a 2D dictionary matching algorithm that that has a linear time complexity and uses sublinear working space.

At first, we focus on the initial rows of the text for which the *LCM* is small enough for calculations to be done in constant time. Let $r$ be the minimum value for which *LCM$_r$* $> m$. Using Algorithm 2, we compute the representative 2D Lyndon word of the first $r$ rows of the text, which we denote by $2D\_LW[1...r]$. This occurs in column $z[r]$ of *LCM-matrix*$[1...r]$, where $z[r]$ represents the value of $z$ after the `for` loop iterates for row $r$ of the input matrix. We then compute the shifted *LWpos* values at columns $z[r]$ and column $z[r] + LCM_r$ for all of the remaining rows. The concatenation of $2D\_LW[1...r]$ with the remaining *LWpos* values gives two particular *LWpos* arrays representing two conjugates of the *LCM-matrix* of the text. The third step is to compare both of these *LWpos* arrays to a trie that was constructed in the preprocessing phase. This trie consists of one *LWpos* array of each pattern with the same 1D names that has $2D\_LW[1...r]$, specifically, the *LWpos* array of each pattern's $z[r]$. We prove in Lemma 8 that this is sufficient to determine which patterns can be consistent with the text block.

The final step is to confirm that the text extends far enough for a pattern to occur, based on the distance between $z[r]$ in the pattern's *LCM-matrix* and the corresponding column of the text block's *LCM-matrix*. We check this in constant time for each pattern in the trie that matched in the previous step. There can be no more than *LCM$_r$* $= O(m)$ such patterns, hence, this completes in $O(m)$ time.

**Lemma 8** *Let $P$ be an $m \times m$ matrix with LWpos array $LWpos_p$ and $T$ an $m \times 3m/2$ matrix with* LWpos *array $LWpos_t$, both $P$ and $T$ have identical 1D names. Furthermore, suppose $P$ and $T$ are classified with identical 2D Lyndon words up* until row $r$, where $LCM_r > m$, and that it occurs in columns $z_p[r]$ and $z_t[r]$ respectively. $P$ occurs within $T$ if and only if $LWpos_p[i]$-$z_p[r] = LWpos_t[i]$-$(z_t[r]+w) \pmod{period[i]}$, for all $r \leq i \leq m$ with $0 \leq z_t[r] + w - z_p[r] \leq m/2 + 1$, where $w = \{0, LCM_r\}$.

**Proof:** Suppose $P$ occurs within $T$. Then, the entire *LCM-matrices* of $P$ and $T$ are conjugate due to the high periodicity of the rows. By Lemma 4, a conjugate of an *LCM-matrix* is recognizable by a shift in the *LWpos* array, with respect to the period size of each row. For the converse, using simple arithmetic, $z_t[r]+w-z_p[r]$ gives the actual constant that shifts $LWpos_t$ to $LWpos_p$. If this constant is smaller than $m/2$, then $P$ occurs within $T$. ∎

This algorithm for 2D dictionary matching incorporates several additional pre-processing steps. We store $LCM[i]$ for $1 \leq i \leq r$ for each 1D name. The *LWpos* arrays for conjugates that begin in column $z[r]$ of the *LCM-matrix* of patterns with the same 1D names are computed and indexed in a compressed trie. The entire pre-processing phase takes $O(dm^2)$ time and uses $O(dm)$ space to index a dictionary of size $O(dm^2)$.

In summary, $O(m^2)$ time verification of a $3m/2 \times 3m/2$ text block is as follows.

For each text block row,

1. Compute $2D\_LW[1...r]$ with Algorithm 2.

2. Compute *LWpos* array in columns $z[r]$ and $z[r] + LCM_r$ of text block's *LCM-matrix*, $LWpos[i]+w-z[r] \pmod{period[i]}$, where $w = \{0, LCM_r\}$.

3. Traverse a compressed trie to compare each of the two *LWpos* arrays from the previous step to shifted *LWpos* arrays of patterns with the same 1D name.

4. For patterns in which the *LWpos* array matches, confirm that the pattern occurrence is close enough to the beginning of the text block's *LCM-matrix*, i.e., $0 \leq z_t[r] + w - z_p[r] \leq m/2 + 1$, with the same value of $w$ as in step 2.

## 6 Conclusion

In this paper we have introduced a new classification scheme for 2D matrices whose rows are all highly periodic. We formulated the 2D Lyndon word and use it as the representative of each consistency class of matrices. We presented several efficient algorithms to compute the 2D Lyndon word that represents a matrix.

Aside from the usefulness of this classification scheme in 2D periodicity, horizontal suffix-prefix matching, and succinct 2D dictionary matching, the new techniques are general and we believe that they will prove useful in other applications for 2D data. In future work, we would like to extend our new techniques to data representing three or more dimensions.

### Acknowledgements

The authors would like to thank Binyomin Balsam for his helpful discussions and his insight into the modular arithmetic solution.